\documentclass[ms,11pt]{egs}       
  \usepackage{times}                   

\usepackage{graphicx}

\printfigures              

\begin{document}
\title{Radio and optical observations of large-scale traveling
ionospheric disturbances during a strong geomagnetic storm of 6-8 April 2000}

%
%
%
%
\author[1,*]{E. L. Afraimovich}
\author[2]{Ya. F. Ashkaliev}
\author[2]{V. M. Aushev}
\author[1]{A. B. Beletsky}
\author[1]{L.A.Leonovich}
\author[1]{O.S.Lesyuta}
\author[1]{A. V. Mikhalev}
\affil[*]{p. o. box 4026, Irkutsk, 664033, Russia\\
          fax: +7 3952 462557; e-mail: afra@iszf.irk.ru\\}
\affil[1]{Institute of Solar-Terrestrial Physics SD RAS}
\author[2]{V. V. Vodyannikov}
\author[2]{A. F. Yakovets}
\affil[2]{Institute of Ionosphere, Almaty, 480020, Kazakhstan}
%
%
\date{Manuscript version from 3 December 2000}

\journal{Journal of Atmospheric and Solar-Terrestrial Physics}       
%
%
\firstauthor{Afraimovich} \proofs{E. L. Afraimovich\\Institute of
Solar-Terrestrial Physics SD RAS\\ Irkutsk, Russia} \offsets{E. L.
Afraimovich\\Institute of Solar-Terrestrial Physics SD RAS\\
Irkutsk, Russia}

\msnumber{12345}

\received{28 August 2000}
\revised{15 September 2000} 
\accepted{3 December 2000}

\runninghead{Smith and Weston: Winning the West}
\firstpage{123}
\pubyear{2001}
\pubvol{25}
\pubnum{2}

\maketitle
\newcommand{\num}{N \stackrel{_O}{_-}}
%
%
%

\begin{abstract}
Basic properties of the mid-latitude large-scale traveling
ionospheric disturbances (LS TIDs) during the maximum phase of a
strong magnetic storm of 6-8 April 2000 are shown. Total electron
content (TEC) variations were studied by using data from GPS
receivers located in Russia and Central Asia. The nightglow
response to this storm at mesopause and termospheric altitudes
was also measured by optical instruments FENIX located at the
observatory of the Institute of Solar-Terrestrial Physics,
($51{.}9^\circ$ N, $103.0^\circ$ E) and MORTI located at the
observatory of the Institute of Ionosphere ($43{.}2^\circ$ N,
$77{.}0^\circ$ E). Observations of the O (557{.}7 nm, 630{.}0 nm,
360-410 nm, and 720-830 nm) emissions originating from
atmospheric layers centered at altitudes of 90 km, 97 km, and 250
km were carried out at Irkutsk and of the $O_{2}$ (866{.}5 nm)
emission originating from an atmospheric layer centered at
altitude of 95 km was carried out at Almaty. Variations of the
$f_{0}F2$ and virtual altitude of the F2 layer were measured at
Almaty as well. An analysis of data was performed for the time
interval 17{.}00-21{.}00 UT comprising a maximum of the $D_{st}$
derivative. Results have shown that the storm-induced solitary
large-scale wave with duration of 1 hour and with the front width
of 5000 km moved equatorward with the velocity of 200 $ms^{-1}$
to a distance of no less than 1000 km. The TEC disturbance,
basically displaying an electron content depression in the
maximum of the F2 region, reveals a good correlation with growing
nightglow emission, the temporal shift between the TEC and
emission variation maxima being different for different altitudes.
\end{abstract}

\section{Introduction}
In the course of strong geomagnetic storms, significant changes
in main structural elements of the magnetosphere and ionosphere
occur. Geophysical manifestations of extremely strong magnetic
storms are of particular interest because these storms take place
relatively rarely (no more than 4 events during an 11-year solar
cycle), and therefore the representative statistics of the whole
complex of interactive processes in the "magnetosphere-
ionosphere" system is lacking.

We have now reached a new quality level in studying these
phenomena because a large number of ionospheric and
magnetospheric parameters are continuously monitored by various
ground-based and space facilities. A new era in the remote
ionospheric monitoring was opened up with the advent of the
Global Positioning System (GPS) now comprising more than 800
world-wide two-frequency GPS receivers whose data are available
through the INTERNET.

Large-scale traveling ionospheric disturbances (LS TIDs) with a
period of 1-2 hours and a wavelength of 1000-2000 km constitute
the most significant mid-latitude consequence of magnetic storms.
Many papers including review papers  (HUNSUCKER, 1982; HOCKE AND
SCHLEGEL, 1996) have been published. LS TIDs are considered to be
a manifestation of internal atmospheric gravity waves (AGWs)
excited by sources in the polar regions of the northern and
southern hemispheres.  Thus, the study of LS TIDs provides
important information on auroral processes under quiet and
disturbed geomagnetic conditions.

AFRAIMOVICH  et al. (2000) were the first to develop  a technique
for determining the LS TIDs parameters based on calculations of
spatial  and temporal gradients of  total electron content (TEC)
measured by three spaced GPS receivers (a GPS  array). This
technique was employed to determine the LS TIDs parameters in the
course of a strong magnetic storm of 25 September 1998. It was
shown that a large-scale solitary wave excited in the auroral
region with a duration of about 1 hour and the front width of
3700 km, at least, traveled equatorward to a distance no less
than 2000-3000 km with the average velocity of about 300 m/sec.

Another interesting consequence of strong magnetic storms is
low-latitude auroras. The global response to the magnetic storm
of the year 1989 was studied by YEH ET AL. (1994). Low-latitude
auroras were observed in the northern and southern hemispheres. A
long-term   electron density depression in the mid-latitude
ionosphere is the most pronounced effect of the a storm. During
the maximum phase of the storm, the zone of disturbances extended
to geomagnetic latitudes of less than $10^\circ$ causing a
temporal depression of the equatorial anomaly.

There appeared many papers on the behavior of  nightglow
emissions of the upper atmosphere during geomagnetic disturbances
(CHAPMAN, 1957; ISHIMOTO ET AL, 1986; TINSLEY, 1979; TORR, 1984).
Several peculiarities in spectra of upper atmosphere emissions at
the middle and low latitudes during strong geomagnetic
perturbations allow them to be classified as "mid- and
low-latitude auroras" (RASSOUL ET AL., 1993) distinguishing from
the "common" aurora at polar latitudes. The differences between
auroras include the appearance of the $N_{2}^{+}$ emission in
bands of the first negative system of mid-latitude spectra, a
significant increase of the atomic oxygen (630{.}0 nm) emission,
and the predominance of emissions of atomic ion lines above these
of molecular bands.

RASSOUL ET AL. (1993) classified several types of low-latitude
auroras in relation to the type of bombarding particles
(electrons, ions, neutral particles), dominating emissions,
localization, and typical temporal scales. Many observations
revealed several types of simultaneously existing auroras caused
by the bombardment of  fast electrons and mixture heavy
particles. At mid-latitudes during moderate geomagnetic
perturbations, 630 nm emission variations with periods ranging
from 0{.}5 to 2 hours were recorded (MISSAWA ET AL., 1984: SAHAL
ET AL, 1988). Mid-latitude auroras occurring in the course of very
strong magnetic storms ($K_{p}$ $\geq$ 8-9,  $D_{st}$ $\geq$ 300
nT) are of the particular interest because the number of
observations with optical instruments was limited.

Although mid-latitude ionospheric storms have been studied during
several decades, there is no complete explanation of their
effects because of the small number of sounding facilities and
their low spatial and temporal resolutions of an ionosonde, a
incoherent scatter radar and optical devices. Moreover, in
contrast to polar latitudes, a small number of observations were
carried out at mid-latitudes simultaneously by radio and optical
techniques which supplement each other because they allow their
shortcomings to be compensated and the reliability of
interpretation of phenomena to be increased.

The objective of this paper is to study the response of the
mid-latitude ionosphere to the strong magnetic storm of 6-8 April
2000 by using data of simultaneous radio and optical observations
in Russia and Central Asia, main attention being paid to LS TIDs
with a characteristic temporal period on the order of 1 hour.
Small-scale disturbance, whose increase in intensity is related
to the shift of the auroral region toward mid-latitudes, will be
the objective of a next study.

Section 2 gives a description of the state of the geomagnetic
field on 6-8 April 2000, and the scheme of the experiment. The
features of LS TIDs obtained on the basis of TEC and optical and
ionosonde data are described in Sections 3 and 4. Section 5 is
devoted to the discussion of the results.

\section{DESCRIPTION OF THE STATE OF THE GEOMAGNETIC FIELD ON
 6-8 APRIL 2000, AND SCHEME OF THE EXPERIMENT}

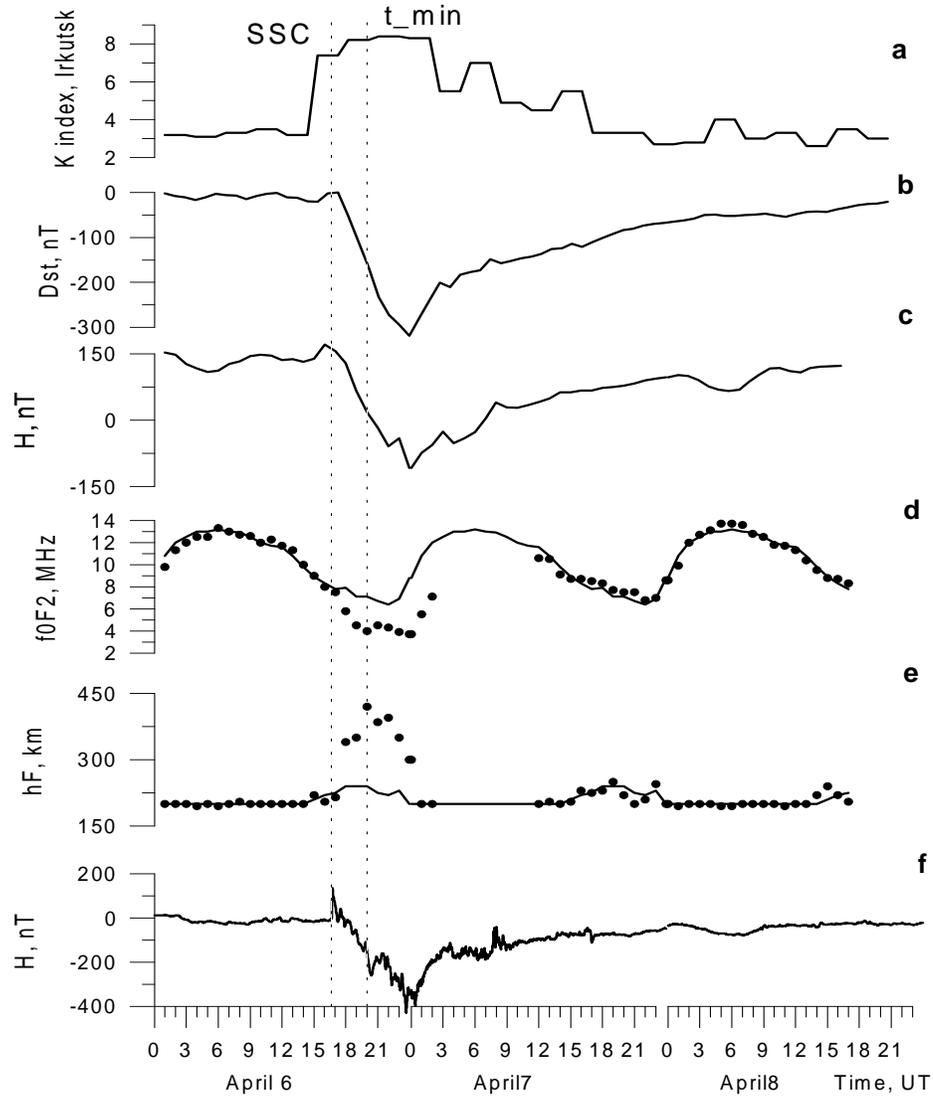
Fig.~1 shows the K-index (a), the $D_{st}$ -variations of the
geomagnetic field, and the variations of the H-component of the
geomagnetic field at Almaty (c) and Irkutsk (f) in the course of
the strong magnetic storm of 6-8 April 2000. This storm was
characterized by a pronounced sudden commencement (SSC) that
started at 1642 UT. At the maximum of the storm, the K-index
achieved the value 8, and a K-index diurnal sum of 48 was
observed. At 1600 UT on 6 April, the $D_{st}$ amplitude increased
fast to 0, but after that it began to decrease, and at 2400 UT it
reached the value -319 nT. After that, the recovery phase
continued into 8 April. In Fig.~1, the dashed vertical lines show
SSC and tmin=20{.}00 UT corresponding to the maximum value of the
time derivative of $D_{st}$  ($dD_{st}/dt$).

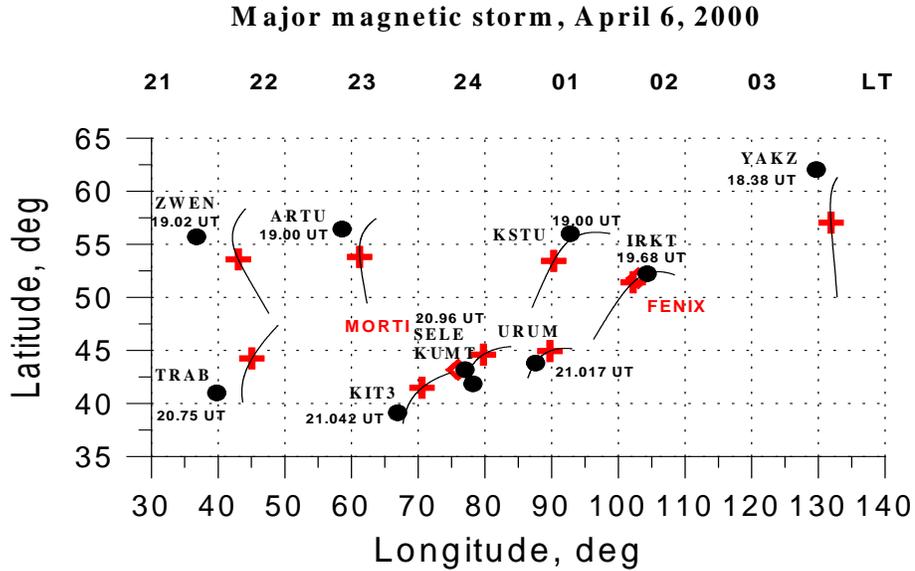
Fig.~2 shows the scheme of the experiment in the geographical
system of coordinates. The positions of the GPS receivers are
denoted by heavy dots, and their names are given. On the upper
scale, the values of the local time (LT) for a certain
longitudinal interval corresponding to the relative time of the
arrival of the LS TID at middle latitudes at 19{.}00 UT are
plotted (Section 3). Diamonds and slant letters show the
positions of optical instruments MORTI (near Almaty and station
SELE) and FENIX (near Irkutsk and station IRKT). Data of an
Almaty standard ionosonde were also used in this paper.

GPS receivers are distributed all over the world with a different
density, and the region considered in this paper comprises only
11 stations whose coordinates are listed in Table~1. Parameters
of LS TIDs are considered to have been determined with a proper
reliability when the distances between GPS receivers exceed the
wavelength of TIDs  (about 1000 km). The array of GPS receivers
used in the experiment satisfied this requirement.

\section{PARAMETERS OF LS TIDs MEASURED BY GPS RECEIVERS AND
THE ALMATY IONOSONDE}

The GPS technique makes it possible to determine the parameters
of TIDs from the phase variations at two carrier frequencies
measured at spaced sites. Methods of  calculating the relative
TEC variations from measurements of the ionosphere-induced change
in the phase path of GPS signals  were described in detail in
several papers  (HOFMANN-WELLENHOF ET AL., 1992; AFRAIMOVICH ET
AL., 1998, 2000). Here we reproduce the resulting expression for
phase measurements:

\begin{equation}
             I_o=\frac{1}{40{.}308}\frac{f^2_1f^2_2}{f^2_1-f^2_2}
                           [(L_1\lambda_1-L_2\lambda_2)+const+nL],
\end{equation}

where $L_1\lambda_1$ and $L_2\lambda_2$ are additional paths of
the radio signal caused by the phase delay in the
ionosphere,~(m); $L_1$ and $L_2$ represent the number of phase
rotations at the frequencies $f_1$ and $f_2$; $\lambda_1$ and
$\lambda_2$ stand for the corresponding wavelengths,~(m); $const$
is the unknown initial phase ambiguity,~(m); and $nL$~ are errors
in determining the phase path,~(m).

For this type of measurements with the sampling rate of 30
seconds, the error of TEC measurements does not exceed $10^14$ $
m^{-2}$, the initial value of TEC being unknown (HOFMANN-WELLENHOF
et al., 1992). This makes it possible to detect irregularities and
waves in the ionosphere over a wide band of amplitudes (up to
$10^{-4}$ of the diurnal TEC variation) and periods (from 24 hours
to 5 min). The TEC unit (TECU), which is equal to $10^{16}$
$m^{-2}$ and commonly, accepted in the art, will be used in the
following.

Initial data to calculate the parameters of LS TIDs were
time-series of TEC at certain sites with corresponding
time-series of the angle of  elevation $\Theta(t)$ and  azimuth
$\alpha(t)$ for the satellite-receiver line calculated by using
software CONVTEC which was able to interpret GPS RINEX-files.
Continuous time-series of $I(t)$ measurements with a duration no
less than 3 hours were chosen for determining the LS TIDs
parameters.

To exclude trends caused by regular changes in ionospheric
density and satellite motion, an hour running average was
subtracted from the TEC time-series. $\Theta(t)$ and $\alpha(t)$
were employed to calculate coordinates of subionospheric points.
To convert the variations of slant TEC to that of vertical TEC,
the a well- known technique (KLOBUCHAR, 1986)

\begin{equation}
I = I_o \times cos \left[arcsin\left(\frac{R_z}{R_z +
h_{max}}cos\theta\right) \right],
\end{equation}

was used, where $R_{z}$ is the Earth's radius, and $h_{max}$ = 300
km is the altitude of subionospheric points (near the F2 layer
maximum).

\subsection{Parameters of large-scale traveling ionospheric
disturbances determined from GPS data}

Fig.~3 and Fig.~4 show the initial $I(t)$ and detrended
time-series $dI(t)$. For the  YAKZ site, only data from the PRN30
satellite for the interval 17{.}00 - 20{.}00 UT were available for
technical reasons.

Almost all GPS records show a gradual decrease of $I(t)$ till a
certain time ($t_{min}$) corresponding to minima (designated by
diamonds in Fig.~3 and 4) in TEC variations,  $t_{min}$  depending
on the latitude of the GPS site. Large fast variations of TEC
occur for some sites after $t_{min}$ has elapsed.

Satellite PRN25 was chosen for all GPS sites analyzed (except
YAKZ) because its minimum elevation angle $\Theta(t)$ exceeded
$45^\circ$ for every station during 19{.}00 - 21{.}00 UT. Thus,
the error of converting the slanting TEC to vertical one caused
by the difference between the actual and spherically-symmetric
spatial TEC distributions was minimized.

In Fig.~2, the solid lines show trajectories of motion of
subionospheric points for satellite PRN25 (PRN30 for YAKZ) at the
altitude of 400 km. Crosses at the trajectories indicate the
positions of subionospheric points at $t_{min}$ corresponding to
minimum TEC (Fig.~3, 4). Near crosses one can find the value of
$t_{min}$ expressed in terms of decimal parts of an hour. For the
subionospheric point of every GPS station, the values of $t_{min}$
and amplitudes ($A_{min}$) expressed in TECU are listed in
Table~1. As is seen from Fig.~2, a minimum $dI$ was first recorded
for the subionospheric point of station YAKZ at $57^\circ$ N
latitude (thin line in Fig.~3c), and, after that, almost
simultaneously it was recorded near $53^\circ$ N latitude at
stations ZWEN, ARTU, KSTU (Fig.~3 a, d; b, e; c, f respectively)
which are extended along the same parallel over the longitudinal
difference of $47^\circ$. Clearly, the ionospheric disturbance had
a wave front with its length exceeding 5000 km. Forty minutes
later this disturbance was recorded at the subionospheric point
for station IRKT at $51{.}5^\circ$ N latitude (Fig.~4a, d). Two
hours later a similar disturbance dI(t) was recorded at a chain
of stations TRAB, SELE (Fig.~4 b, e), KITS, KUMT and URUM (Fig.~4
c, f) at  $40^\circ$ - $45^\circ$ N latitudes.

By using Table~1 and Fig.~3 and 4, one can study the evolution of
the disturbance as it travels equatorward. The amplitude
($A_{min}$) of the disturbance $dI(t)$ decreases from 5 TECU at
the northern chain of stations to 1 TECU at the southern chain.
Moreover, large fast variations $dI(t)$ typical of the
high-latitude ionosphere were recorded after passing a minimum
$dI(t)$ at the northern chain of stations. The same result was
obtained by AFRAIMOVICH ET AL. (2000). These variations are
noticeably less at the southern chain of stations.

These features of the $dI(t)$ variations seem to be accounted for
the fact that at about 19{.}00 UT satellite-receiver lines for the
northern chain of stations (Fig.~2)  crossed the southern boundary
of the auroral zone moving southward. If the front of the
disturbance has traveled with a constant velocity, then a delay
on the order of 2 hours between the times of the rise of the
disturbances at the northern and southern chains of stations
corresponds to the southward velocity of about 200 m/sec.
However, by using the technique of AFRAIMOVICH ET AL. (2000), one
can obtain the velocity of about 500 m/sec, and the wave vector
directed southeastward. By means of a chain of ionosondes, MAEDA
AND HANDA (1980) and Whalen (1987) obtained similar results. The
extension of the disturbance front on the order of 5000 km
obtained in this study is consistent with results reported by
WHALEN (1987),  HAJKOWICZ,  AND HUNSUCKER (1987).

So, in the main phase of the strong magnetic storm, a significant
descent on the order of 15-20 TECU was observed at the northern
chain of GPS sites, including ZWEN, ARTU, KSTU and YAKZ.
HUNSUCKER (1982) and BALHAZOR AND MOFFETT (1999) showed that a
large area of the polar atmosphere leaving abruptly the state of
equilibrium must become the source of LS TIDs traveling
equatorward. It is necessary to point out that this period of
time was characterized by a maximum time derivative $D_{st}$
(Fig.~1c) which is consistent with the conclusion of HO ET AL.
(1998).

\subsection{Variations of the critical frequency and virtual
altitude of the F2 layer over Almaty}

Fig.~1d shows the variations of the critical frequency of the F2
layer ($f_{0}F2$) (large dots). The solid line represents the
$f_{0}F2$ current median defined from three- month data (WRENN ET
AL., 1987). The day before the magnetic storm was magnetically
quiet, and $f_{0}F2$(t) was close to its median values. The main
phase of the magnetic storm occurred at local night at Almaty. A
decrease of $f_{0}F2$ relative to the median, reflecting a
depression of the electron content in the F2 layer maximum, began
soon after the beginning of the storm and a maximum difference
between the current $f_{0}F2$ and median values took place at the
period of a maximum time derivative $dD_{st}/dt$. The growth of
$f_{0}F2$ after the solar ionizing agent appeared is delayed with
respect to the median by two hours. Unfortunately, during 7 hours
after 02{.}00 UT the ionosonde did not operate for technical
reasons; therefore, the time at which f0F2 approached median
values was uncertain.

Variations of the virtual altitudes $h_{'}F$ are plotted in
Fig.~1e with large dots. The solid line shows the behavior of the
$h_{'}F$ current median. It is seen from Fig.~1e that an abrupt
increase of $h_{'}F$ with respect to median values began
simultaneously with the decrease of $f_{0}F2$. The approach to
the median values occurred at 01{.}00 UT. At that time $f_{0}F2$
was equal to 6{.}0 MHz, and the median value was 11 MHz. The large
decrease of $f_{0}F2$ in the course of the main phase of the
magnetic storm is consistent with the large negative disturbance
of TEC recorded by GPS stations (Fig.~4).

Thus, this magnetic storm was accompanied by a very large
decrease of the electron content in the maximum of  the F2 layer,
and by a significant increase of its virtual altitude.

\section{OPTICAL OBSERVATIONS OF NIGHTGLOW EMISSIONS}
\subsection{Large-scale disturbances of nightglow
emissions over Irkutsk }

An optical facility FENIX is settled at Geophysical observatory
attached to Institute of Solar-Terrestrial physics at 100 km from
Irkutsk city ($51{.}9^\circ$ N, $103{.}0^\circ$ E ; geomagnetic
latitude is $41{.}0^\circ$ N, L=2). It includes a four-channel
zenith photometer and a high sensitive TV-system comprising an
image amplifier and a CCD (charge coupled device) array.

Following observations of nightglow emissions were made: the OI
(557{.}7 nm) emission which originates from a layer centered at 97
km and with boundaries at altitudes of 85 km and 115 km, the OI
(630{.}0 nm) emission which originates from a layer centered at
250-270 km and with boundaries at altitudes of 160-300 km, the
$O_{2}$ (360-410 nm) emission which originates from a layer
centered at 97 km, and the OH emission which originates from a
layer centered at 85-90 km and with boundaries at altitudes of
75-115 km.

Emission lines of 557{.}7 and 630 nm were recorded by using narrow
band (1-2 nm halfwidth) sweeping interference filters. Records of
360-410 nm and 720-830 emissions were obtained by using broad
band absorption filters. The angle of view of instruments was
about 4-5. The instruments were absolutely calibrated
periodically by using the standard stars and were controlled by
using reference lamps in periods between absolute calibrations.
The software provided records of the emission rate with an
exposure time of 12 sec, but when the emission rate exceeded the
predetermined level, a shorter exposure time of 8 ms was used. In
the course of the magnetic storm of 6 April 2000, the records
were taken only by the four-channel zenith photometer.

Variations of the nightglow emission rate, derived from the zenith
photometer observations on 6 April 2000, are plotted in Fig.~5a.
The main feature of these variations is a significant increase of
the OI (630{.}0 nm) emission in the second half of the night
(Fig.~5a, line 1) by a factor of 20 compared with values observed
near midnight and the last geomagnetic quiet night of 5 April
2000 (Fig.~5a, line 3). It can be seen that the OI (630{.}0 nm)
emission grew after 16{.}00 UT till sunrise when the observations
were completed, and periodical variations were superimposed on
the gradual growth. The OI (557{.}7) emission variations
(Fig.~5a, line 2) revealed a small disturbance near 17{.}00 UT
coinciding with a similar disturbance in the OI (630{.}0 nm)
emission, and an abrupt rise (35$\%$) coinciding with the first
phase of the OI (630{.}0 nm) emission rate maximum increase.

A gradual increase of $O_{2}$ (360-410 nm) emission, not typical
of the quiet geomagnetic state (Fig.~5b, line 6 for the previous
night of 5 April 2000) (Fig.~5b, line 4) was observed with
superimposed irregular short-period variations, to begin at
17{.}00 UT. The 720-830 nm emission (Fig.~5b, line 5) revealed a
gradual decrease from 14{.}00 till 16{.}00 UT and ended at 16{.}00
UT when the commencement of the geomagnetic storm occurred.

It is of interest to compare these data with variations of TEC
measured at the nearest GPS station IRKT. Records of the 630 nm
and 577{.}7 nm emission lines  ($A(t)$) and ($B(t)$) filtered out
from the initial data (Fig.~5a, lines 1 and 2 respectively) as it
was done with TEC $I(t)$, are plotted in Fig.~4d. It is seen that
there is a good correlation between these variations, and the
emission variations are in an opposite phase compared with TEC
variations. The possible reasons for this behavior are discussed
in part 6.

\subsection{Large-scale disturbances observed by the MORTY
instrument}

The MORTI (Mesopause Rotational Temperature Imager) instrument is
installed near Almaty in the mountains at 2800 m above sea level
($43{.}05^\circ$ N, $76{.}97^\circ$ E). MORTI provides
information for obtaining the rotational temperature and emission
rate of the $O_{2}$ atmospheric (0-1) nightglow layer centered at
95 km (WIENS ET AL., 1991). The instrument comprises a conical
mirror to receive the light from a full circle, a narrow band
(0{.}27 nm halfwidth) interference filter centered at 867{.}6 nm,
an imaging lens to focus the spectrum, and a CCD camera to record
the spectrum. The emission rate and temperature observations have
the precision of $\pm2$ $\%$ and $\pm2$ K, respectively, when the
exposure time is 5 min.

Variations of the $O_{2}$ (867{.}6 nm) emission on 6 April 2000
are plotted by dots in Fig.~6a, and detrended variations are shown
in Fig.~6b. The quadratic trend calculated by the  least square
technique (Fig.6a,  solid line) was subtracted from the initial
record to obtain the detrended variations. It is seen that large
variations in the $O_{2}$ (867{.}6 nm) emission were observed even
before the time of the arrival of the solitary wave caused by the
magnetic storm (this time is known from the GPS data). The period
of background variations was about 2 hours, so the solitary wave
excited by the magnetic storm can be distinguished because its
period was about 1 hour. In order to reduce the effect of
two-hour background variations the high pass filter realized by
the hour running window was applied to the record in Fig.~6b.
Filtered variations of the $O_{2}$ (867{.}6 nm) emission are
plotted in Fig.~4e together with variations of TEC $I(t)$ for the
GPS station SELE. It is seen from Fig.~4e that there is a good
correlation between the variations from the GPS and MORTI data.
By comparing Fig.~4d and Fig.~4e, it becomes evident that the
phase delays between the variations of the GPS and MORTI data are
different for Irkutsk and Almaty. This difference is explicable
by the different altitudes of the nightglow emissions, and the
different types of optical instruments.

\section{DISCUSSION}

The parameters of the disturbances considered in this study are
affected by various factors. These are the localization and
dynamics of the main ionospheric trough (MIT), the plasmasphere
and the plasmapause, the zone of precipitating particles (auroral
oval), and magnetospheric-ionospheric currents. During periods of
great geomagnetic activity, the plasmapause boundary may approach
extreme values of L in the range of 1{.}7-2{.}5 (KHOROSHEVA,
1987), the position of MIT - about $43^\circ$ N of invariant
latitude (ANNAKULIEV ET AL., 1997), and the auroral oval
boundaries - about $48^\circ$ N corrected geomagnetic latitudes
(http://www.sec.noaa.gov/Aurora/index.html). For Irkutsk this
latitude is equal to $47^\circ$ N. According to the data from
satellite NOAA-15 (http://sec.noaa.gov/\\pmap/pmapN.html) near
sunrise of 7 April 2000, the auroral oval boundary (on the level
of 0{.}1 $erg/cm^{2}s$) approached $56-58^\circ$ N geographical
latitudes in the longitude sector under consideration. Taking into
consideration that the width of MIT is of several degrees, it is
possible to consider that, at least, the southern boundary of MIT
approached the latitude of $52^\circ$ N in Eastern Siberia where
optical observations were carried out. These facts suggest that
in the course of the magnetic storm considered, elements of the
subauroral and even auroral ionosphere were observed at latitudes
of Irkutsk. This conclusion is supported by some features of
optical observations. The signal gain in the spectral band of
360-410 nm after 17{.}00 UT may be the result of a rise of the
$N_{2}^{+}$ (ING) emission  with a wavelength of 391{.}4 nm which
is usually observed in auroras after the ionization of $N_{2}$ by
precipitating electrons or  precipitation of energetic atoms or
ions (ISHIMOTO et al., 1986; TINSLEY et al., 1984).

Beside the known fact of the increase of the 630{.}0 nm emission
in midlatitudes auroras, there is a maximum in the 630{.}0 nm
emission at about 19{.}30 UT, which corresponds to a minimum TEC
at almost the same place (Fig.~4d). The same behavior of the
630{.}0 nm emission was observed above Irkutsk in the course of
the strong magnetic storm of 24-25 March 1991 (MIKHALEV, 1997).
If  a minimum in TEC variations is related to the passage of  MIT
during the increase of the ring current in the start phase of the
magnetic storm, then it is possible to conclude that the maximum
630{.}0 nm emission develops in the region of MIT where stable
auroral red (SAR) arcs occur (REES and ROBLE, 1975).

In connection with the issue discussed, the results of LOBZIN AND
PAVLOV (1998) on SAR arcs above North America are of certain
interest.  In the cited paper, it was pointed out that the pale
glow in the form of SAR arcs might become the phenomenon of
low-latitude aurora during strong magnetic disturbances. During
strong magnetic storms ($D_{st}$ $>$ -200 nT) the 630 nm emission
rate from the SAR arcs approaches 2{.}2 - 4{.}0 kR. The 630 nm
emission rate recorded during the magnetic storm considered above
Irkutsk is in this emission range. Identifications of the SAR
arcs and midlatitudes lights were carried out at other papers (for
example, KHOROSHEVA, 1987).

For the SAR arcs, an increase of populating $^{1}D$ neutral oxygen
originates from growing electron temperature at altitudes of the
F2-layer and outer ionosphere caused by an increase of the heat
flow from the plasmasphere where an interchange of energy between
the heat plasma and ring current occurs. The different pattern of
the change of short-period variations in the nightglow emissions
before and after the TEC minimum must be emphasized where the
amplitude of these variations after the TEC minimum is sharply
increased. Besides, they correlate with short-period variations
of TEC  (Fig.~3 and Fig.~4 a, d). The rise of the short-period
variations appears to be caused by precipitation of energetic
particles.

According to BURNS et al. (1991), TEC is directly proportional to
the ratio $O$ / ($O_{2}$ + $N_{2}$), therefore when the $N_{2}$
increases the local and integral electron contents decrease, and
this fact is consistent with results obtained (Fig.~4).

\section{CONCLUSION}

Thus, in the course of the magnetic storm of 6 April 2000, two
types of disturbances were observed. The first one has features
of a solitary wave with a period of about 1 hour, and was
interpreted as an LS TID originating in the polar latitudes. The
second one included short-period variations probably related to
the particle precipitation. Alternatively, the wave-like
disturbance may be interpreted as a motion of structural element
projections of the disturbed ionosphere (the plasmapause and MIT)
on to the region of midlatitudes during the start stage of the
magnetic storm. In this case, during a strong magnetic storm the
elements of the subauroral ionosphere were observed at the
latitudes of Irkutsk.

An analysis of the data has shown that being originated the
auroral disturbance induced LS solitary wave with a period of
about 1 hour and the front width no less 5000 km traveled
equatorward to a distance no less than 1000 km with the average
velocity of about 200 m/s. The TEC disturbance, showing mainly a
decrease of the electron content in the vicinity of the F2-layer
maximum, correlates with an increase of the emission rate in the
optical band, with the temporal shift being different for
different ionospheric altitudes.

At Almaty the magnetic storm was followed by an extremely large
decrease of the electron content in the F2-layer maximum which
caused a decrease of the nighttime critical frequencies by 4-5
MHz. The decrease of the electron content was accompanied by a
significant increase of the F2-layer virtual height.

\begin{acknowledgements}
We are grateful to E.A. Ponomarev and A.V. Tashchilin for their
encouraging interest in this study, and for active discussions.
This work was done with support of INTAS grant $\num$ 99-1186,
Leading Scientific Schools of Russian Federation grant $\num$
00-15-98509 and Russian Foundation for Basic Research grant $\num$
99-05-64753.
\end{acknowledgements}

\end{document}